\documentclass[manuscript,nonacm]{acmart}
\usepackage{booktabs}
\usepackage{pifont}
\usepackage{array}
\AtBeginDocument{%
  }

\setcopyright{none}

\begin{document}

\title{Assessing Data Literacy in K--12 Education: Challenges and Opportunities}

\author{Annabel Goldman}
\affiliation{%
  \institution{Northwestern University}
  \city{Evanston}
  \state{Illinois}
  \country{United States of America}}
\email{annabelgoldman2025@u.northwestern.edu}
\orcid{0009-0004-5600-725X}

\author{Yuan Cui}
\affiliation{%
  \institution{Northwestern University}
  \city{Evanston}
  \state{Illinois}
  \country{United States of America}}
\email{yuancui2025@u.northwestern.edu}
\orcid{0000-0002-2681-6441}

\author{Matthew Kay}
\email{matthew.kay@u.northwestern.edu}
\orcid{0000-0001-9446-0419}
\affiliation{%
  \institution{Northwestern University}
  \city{Evanston}
  \state{Illinois}
  \country{United States of America}
}

\renewcommand{\shortauthors}{Goldman et al.}


\begin{abstract}
Data literacy has become a key learning objective in K--12 education, but it remains an ambiguous concept as teachers interpret it differently. When creating assessments, teachers turn broad ideas about ``working with data'' into concrete decisions about what materials to include. Since working with data visualizations is a core component of data literacy, teachers’ decisions about how to include them on assessments offer insight into how they interpret data literacy more broadly. Drawing on interviews with 13 teachers, we identify four challenges in enacting data literacy in assessments: (1) conceptual ambiguity between data visualization and data literacy, (2) tradeoffs between using real-world or synthetic data, (3) difficulty finding and adapting domain-appropriate visual representations and data visualizations, and (4) balancing assessing data literacy and domain-specific learning goals. Drawing on lessons from data visualization, human–computer interaction, and the learning sciences, we discuss opportunities to better support teachers in assessing data literacy.

\end{abstract}
\maketitle

\section{Introduction}
As data becomes central to everyday reasoning, K--12 instruction is increasingly emphasizing data literacy as a key learning objective \cite{DS4E, Ravi2024, Schanzer2022}. In response, standards and policy documents call on teachers to develop students’ data literacy skills \cite{NGSS, DS4E}. However, despite this increased emphasis, data literacy remains a vague concept in classrooms as teachers interpret and apply it differently in their instructional practices.

Assessment authoring provides a useful lens for examining how teachers think about data literacy. As teachers author assessments, they must translate broad ideas about working with data into concrete decisions about assessment design, requiring them to decide what data students will encounter, how it will be represented, and what kinds of reasoning will count as evidence of understanding. These decisions are made under time and tooling constraints and they shape which aspects of data literacy are emphasized in classroom practice.

% Assessment authoring therefore provides a useful lens for examining how teachers work with data literacy.

In a prior codesign study aimed at developing an AI-supported assessment authoring tool, we asked teachers about their use of data visualizations in classroom assessments. Although we did not directly ask about “data literacy” as a formal construct, working with data visualizations is widely recognized as a key aspect of data literacy \cite{Ravi2024, Cruickshank2024}. As a result, teachers’ decisions about when and how to use visualizations in assessments provide a window into how they understand data literacy more broadly. 

% Working with data visualizations is a key aspect of data literacy, spanning practices such as interpreting tables, reasoning from maps, and using representations to support claims \cite{Wilkinson1999, Bertin1983}. Because data visualizations often serve as a visible, concrete instantiation of \textit{``working with data,''} teachers’ decisions about when and how to include them on assessments provide a window into how they understand data literacy more broadly. 

Through interviews with 13 teachers, we identify four challenges in assessing data literacy in K--12 education:

\begin{itemize}
    \item Challenge 1: Conceptual ambiguity between data visualization and data literacy
    \item Challenge 2: Tradeoffs between using real-world or synthetic data
    \item Challenge 3: Difficulty finding and adapting domain-appropriate visual representations and data visualizations
    \item Challenge 4: Balancing assessing data literacy and domain-specific learning goals 
\end{itemize}

Drawing on insights from data literacy, visualization, human–computer interaction, and the learning sciences, we discuss opportunities to better support teachers in assessing data literacy within K--12 classrooms. These include providing professional development and lesson plans that distinguish data visualization from broader data reasoning practices; lowering the overhead of cleaning and modifying real-world data; supporting discovery and lightweight adaptation of domain-appropriate visual representations; and structuring assessment tasks so that data literacy and domain knowledge can be assessed in tandem without introducing prohibitive grading burdens.

\section{Methods}

\aboverulesep=0ex
\belowrulesep=0ex
\setlength\arrayrulewidth{.2pt}
\bgroup
\begin{table}[!t]
\centering
\caption[]{\textbf{Teachers' Demographics.} Our teachers represent a wide range of teaching experience, school types, and subject areas.
}
\small
% \fontsize{7.5pt}{10pt}\selectfont
\vspace*{-8pt}
\def\arraystretch{1}% 
% \sffamily
% \color{darkgray}
\label{tab:codesign_partners}
% \sffamily\selectfont
\begin{tabular}{>{\arraybackslash}m{0.04\textwidth}
                m{0.04\textwidth}
                m{0.04\textwidth}
                m{0.04\textwidth}
                m{0.06\textwidth}
                m{0.25\textwidth}}
\midrule\\[-6pt]
\textcolor{black}{\textbf{Id}}   & 
\textcolor{black}{\textbf{Sex}} & 
\textcolor{black}{\textbf{Years}} & 
\textcolor{black}{\textbf{State}}  & 
\textcolor{black}{\textbf{School}}  & 
\textcolor{black}{\textbf{Subjects}} \\ [2pt]
{P1}  & F & 22 & OH & Private & Algebra, Pre-Calculus, Calculus   \\
{P2}  & M & 26 & OH & Private & History, Geography, Economics    \\
{P3}  & M & 28 & OH & Private & Biology, Ecology          \\
{P4}  & M & 27 & IL & Public  & Chemistry          \\
{P5}  & M & 23 & IL & Public  & Physics, Biology          \\
{P6}  & F & 28 & IL & Public  & Biology          \\
{P7}  & M & 22 & IL & Public  & Biology          \\
{P8}  & F & 9  & IL & Public  & Biology, Health Careers          \\
{P9}  & M & 6  & OH & Public  & Pre-Calculus, Calculus              \\
{P10} & F & 2  & OH & Public  & Geometry, Algebra              \\
{P11} & F & 6  & OH & Public  & Algebra, Pre-Calculus             \\
{P12} & M & 11 & OH & Public  & Algebra, Computer Science              \\
{P13} & M & 26 & NY & Private & History    \\
\midrule
\end{tabular}
\end{table}
\egroup

We use data from a previous study that we conducted, originally aimed at developing an \textsc{AI}-powered assessment authoring tool \cite{Cui2026}. We engaged in a 7-month codesign study with 13 high school teachers (Table~\ref{tab:codesign_partners}) consisting of multiple rounds of semi-structured interviews. In our initial interviews, we asked teachers about their existing assessment authoring practices, including how they design and evaluate assessments. 
% These interviews were intentionally broad in scope to surface teachers’ current workflows, constraints, and decision-making processes before introducing any tool concepts. 
As part of this discussion, we asked teachers about their use of data visualizations in assessments. Example questions from this interview protocol included:
\begin{itemize}
    \item How (and how often) do you typically evaluate students?
    \item What challenges do you currently face when designing and creating questions?
    \item Do you use data visualizations in your class?
    \item Do you ask questions about data visualizations?
    \item Is it important that the questions are factual (based on real datasets or scientific facts)? Is fictional context okay?
\end{itemize}

These questions allowed teachers to describe how they include data visualizations and broader data literacy practices on assessments. Importantly, the interviews were not framed around formal definitions of \textit{``data visualization''} or \textit{``data literacy,''} enabling teachers to articulate their own interpretations and priorities. Using thematic analysis, we analyzed the same interview transcripts and identified themes that characterize how teachers think about data literacy and how those viewpoints shape assessment practices.

% All interviews were audio-recorded, transcribed, and analyzed using thematic analysis. 

\section{Teacher Perspectives on Defining and Assessing Data Literacy}
\label{sec:findings}

\subsection{Overlapping Definitions of Data Visualization and Data Literacy}

Teachers' responses to \emph{``do you use data visualizations in your assessments?''} revealed differences in their definitions of visualizations. Teachers used this phrase to describe a wide range of artifacts and practices, including traditional data visualizations, visual representations that contain little or no numerical data, and data literacy tasks without visual representations.

\subsubsection{Teachers Use ``Data Visualization'' to Refer to Many Kinds of Representations}
\label{subsubsec:kinds_of_representations}
Teachers used ``data visualization'' to refer to a broader set of visual artifacts than conventional definitions would suggest. In some cases, teachers’ descriptions aligned closely with conventional definitions of data visualization \cite{Bertin1983, Wilkinson1999}. For example, P1 described using \textit{``dot plots, histograms, [and] box plots''} on assessments. Similarly, P11 described creating \textit{``bar chart[s] or pie chart[s] on like Excel.''} 
% These assessment materials are closely alligned with conventional definitions of data visualization, where numerical values are encoded through graphical marks .

Teachers also described using visual representations that encode little to no data. P5, a physics teacher, explained that for \textit{``a lot of physics questions, you often want a diagram to go along with the question.''} P2, a history teacher, described using infographics as a core part of assessment and instruction. 
% explaining that \textit{``a lot of multiple choice questions in human geography [use] data visualization[s]''} (p2). 
P8, a science teacher, similarly described using \textit{``maps that show specific data,''} including \textit{``a map from 2008 that shows kelp forests, and then in 2019 showing zero kelp forests.''} In these examples, teachers referred to a wide range of visual representations as data visualizations, even when they encode little to no quantitative data.

% Finally, teachers’ definitions also included data literacy tasks without a visual encoding (P4, P1, P8). We examine this subset of responses in detail in Sec. \ref{subsubsec:data_reasoning_practices}. These vastly different responses show that teachers use \textit{``data visualization''} more broadly than definitions that limit the term to charts and graphs. This broad definition may arise for several practical reasons: these materials play similar roles on assessments, are created and sourced through similar workflows, and are not clearly distinguished from other assessment materials in teachers’ everyday language.

\subsubsection{Visualization Talk Often Included Broader Data Literacy Skills}
\label{subsubsec:data_reasoning_practices}

When asked how they include data visualizations on assessments, some teachers instead described tasks where students work with data without any visual representations. For example, P4 said that they often use \textit{``a data table''} to provide \textit{``the values [their students] need in order to solve [a particular question].''} Similarly, P1 described \textit{``pulling numbers''} and \textit{``adjusting them''} so that students could identify outliers or a particular value from the data itself. 

These overlapping understandings may be because teachers are unsure how to distinguish data visualization from data reasoning, see the two as inseparable, or use them in similar ways when designing assessments. This points to \textbf{Challenge 1: Conceptual ambiguity between data visualization and data literacy}. Because the boundary between these constructs is unclear, assessment tasks intended to target one skill may implicitly assess another, and it becomes difficult to specify what kind of student work a ``data visualization'' question is meant to elicit.

\subsection{Sourcing, Creating, and Adapting Data, Data Visualizations, and Visual Representations}
When designing assessments, teachers need to source, create, or adapt data, visualizations, and other visual representations. These decisions reflect how teachers’ broad interpretations of data literacy are translated into concrete assessment materials and are shaped by trade-offs between speed, quality, and alignment with instructional goals. \looseness=-1

\subsubsection{Data Source Choices Shift Based on Instructional Goals}
\label{subsubsec:data_source_choices}
Teachers' instructional goals strongly influenced whether they chose to use real or synthetic data in their assessments. Several teachers described creating or modifying data to surface specific concepts or address common student misunderstandings. For example, P11 explained that while they \textit{``try to make [the data] factual,''} this priority changes when the goal is to assess a particular skill. When discussing student difficulties with range and outliers, they described intentionally creating datasets with low outliers, stating, \textit{``I'll probably make something that's not accurate [to] show them a chart that looks like that.''} P1 described a similar tension, explaining that while they do not want to \textit{``fabricate too much data,''} they still modify datasets to ensure that they contain features such as a desired number of outliers or a particular median. These accounts suggest that synthetic or modified data offers teachers greater control, allowing them to align assessments more directly with instructional goals.

At the same time, some teachers emphasized the value of real data, particularly for supporting the assessment of broader data literacy skills. P5 showed enthusiasm for a dataset drawn from an actual field study because the data violated students' expectations of clean patterns: \textit{``these numbers don't even come close to the 10\% rule that students are gonna expect the data to fit. This is wonderful.''} In this case, real data was valued because it required students to deal with variability and uncertainty. This leads to \textbf{Challenge 2: Tradeoffs between using real-world or synthetic data}, where teachers must choose between using ``clean'' data that target a specific skill and ``messy'' data that better represent reality.

\subsubsection{Hand-Created Visual Representations Offer Control but Are Perceived as Messy or Unprofessional}
\label{subsubsec:hand_created_visuals}
% Teachers’ accounts also included how visual representations and data visualizations are produced for assessments. 
Some teachers create visual representations for their assessments by hand because this approach offers flexibility and control, but expressed concerns about quality and professionalism. P11 noted that while it is possible to create data visualizations by hand, \textit{``normally it looks like really bad.''} Similarly, P5 had previously hand-drawn diagrams, but stopped after receiving feedback that their drawings appeared \textit{``unprofessional.''} 
% despite believing that they were \textit{``carefully drawn.''} 
They also reflected on their own skill, stating \textit{``I'm just not very good at making diagrams.''} As a result, P5 abandoned hand-created visuals in favor of finding pre-existing diagrams online.

\subsubsection{Specialized Tools Promise Quality but Carry High Learning and Time Costs}
\label{subsubsec:specialized_tools}

Some teachers described using specialized tools to generate data or visualizations, but these tools were often used sparingly due to learning and time costs. For example, P13 explained that when making graphs, they typically \textit{``make it in some graphing tool like Desmos''}, then take a screenshot and paste it into a Word document. This approach allowed them to produce precise graphs, but required additional steps to integrate the result into an assessment.

Some teachers also avoid specialized software. P4 noted that, although they had \textit{``access to organic [chemistry] software,''} they did not use it because the efficiency of using pre-made graphics outweighs the quality benefits of using specialized software. In contrast, P8 described using \textit{``SAGE Modeler''} to generate interaction charts because they can produce artifacts closely aligned with instructional goals. These accounts suggest that, while specialized tools can support more sophisticated visual representations, the time and learning costs limit teachers' adoption.

\subsubsection{Pre-Made Data Visualizations and Visual Representations Are Fast but Difficult to Customize and Hard to Find at the Right Quality}
\label{subsubsec:pre_made_visuals}

Many teachers relied on pre-made visualizations drawn from textbooks, online searches, or test banks, largely because these materials can be accessed quickly. As P3 explained, it is \textit{``very common''} for them to \textit{``go out, [grab] something from the net, modify it, and then [copy] and paste it''} into their assessments. They described taking snippets from digital textbooks or online sources and making small edits using Microsoft Paint, noting that creating visuals from scratch was \textit{``not very common''} for them. P11 similarly described turning to existing charts when time is limited, explaining that \textit{``if I'm crunched for time, I'll just Google image [search] and look at just like basic charts and then I'll like make [my] problem around just like the chart that I have.''} In this case, creating a new representation takes more time than revising the assessment, so P11 adjusted the question to fit an existing visualization. When P11 creates more customized visualizations, the process becomes time-consuming and frustrating. They copy screenshots into Google Drawings and \textit{``type over like the numbers [to] make it [their] own,''} summarizing the workflow as \textit{``convoluted,''} even though it is \textit{``the best way [they] could do it.''}

% P4 described frequently relying on online image searches for pre-created graphical representations rather than generating the data and representation themselves, saying: \textit{``I'll just, [type in] I want isobutane and boom, [I'll] get an image [and] dump it [into] my test.''} P4's emphasis on immediacy and effort: \textit{``I'll just dump it in''} highlights speed and convenience as key factors in their choice to use pre-made visualizations. 

Quality was another recurring concern. P5 noted that they had never encountered a test bank with high-quality questions involving physics diagrams.
% noted that they had \textit{``never seen a good test bank for physics,''} explaining that physics questions often require diagrams that are difficult to find at an appropriate level. They described many available resources as \textit{``really, really dumbed down,''} leading them to rely on a small number of external websites, such as Problem Attic. 
P6 echoed this concern, noting that it is \textit{``just hard to find data and visualizations for the level I'm teaching.''} This introduces \textbf{Challenge 3: Difficulty finding and adapting domain-appropriate visual representations and data visualizations}, where limits on time and tools strongly shape what visual representations and data visualizations are used in assessments. 

\subsection{Assessing and Teaching Data Literacy in the K--12 Classroom}

Instead of being assessed as a separate skill, data literacy is often implicitly assessed within subject-specific tasks. When assessed explicitly, new tensions arise around which domain-specific concepts are de-emphasized to make room.\looseness=-1 
% This section examines when data literacy remains implicit and when it is made explicit through standards-driven approaches.

\subsubsection{Data Literacy Is Often Assessed Implicitly}
\label{subsubsec:implicit_data_literacy}
Teachers described assessment tasks where students work with data even when \textit{``data literacy''} or \textit{``visualization literacy''} is not explicitly identified as an assessment goal. For example, P8 described using \textit{``interaction chart[s]''} that students are expected to \textit{``add to''} during an assessment. P4 similarly expected students to \textit{``know how to manipulate [data]''} that they provide in a chemistry test. 
% P1 also described assessments in which students are asked to create \textit{``dot plots or histograms''} from provided data.
These examples show that students practice data literacy through routine assessment work: they extend structured representations, locate and use relevant values, and sometimes convert raw data into a display, all in service of completing a domain task \cite{Yousef2021, Debruyne2021}. 
% These sorts of tasks on assessments often implicitly assess data literacy even when it isn't an explicit instructional goal.

\subsubsection{Three-Dimensional Assessments Make Data Literacy Explicit}
\label{subsubsec:3d_assessments}

Several teachers mentioned a new assessment approach named \textit{``three-dimensional assessments.''} These assessments aim to assess both students' understanding of domain-specific content and their ability to work with data as a cross-cutting skill \cite{NGSS}. As P7 explained, three-dimensional questions need to \textit{``assess whether someone has mastered the domain knowledge [as well as] their ability to understand data.''} P5 explained that students are presented with a \textit{``phenomenological''} question and \textit{``more than just the data that's necessary''} for answering that question. Then students are asked to decide \textit{``which data they're going [to] analyze,''} and \textit{``[provide] evidence to support the claim or [to] refute it.''} 

Several teachers also noted that this approach shifts emphasis away from detailed factual knowledge toward data literacy skills. P6 described how these assessments focus on \textit{``thinking like a scientist, not necessarily knowing the facts of photosynthesis.''} P7 emphasized that the standards themselves explicitly de-emphasize certain content details, noting that these standards suggest students \textit{``don't need to know the biochemistry''} so long as they can \textit{``process data about it''}. Three-dimensional assessments are more explicitly designed to assess students’ data literacy.

\subsubsection{Balancing Instructional Goals and Assessment Feasibility}
\label{subsubsec:feasibility_tensions}

While teachers recognized the value of making data literacy explicit, several expressed ambivalence about how this emphasis affects domain-specific learning goals. In particular, teachers worried that focusing too heavily on cross-cutting skills could come at the expense of content knowledge they viewed as foundational. For example, P6 said: \textit{``I still think kids need to know like basic biology. They don't all have to analyze graphs all day.''} P7 described recognizing that students might not really \textit{“need to know”} specific domain content, while still feeling \textit{``defensive''} about letting go of that knowledge in favor of more cross-cutting skills.

In addition, teachers highlighted practical challenges that limit their ability to adopt three-dimensional assessments. P5 in particular pointed to grading as the main barrier. They explained, \textit{``generating those data is pretty easy''} and \textit{``coming up with the questions is usually pretty easy,''} but \textit{``grading it is just awful.''} Because these assessments are open-ended and often require students to justify claims using evidence, evaluating responses takes substantial time.

% This grading burden also shaped how teachers described using three-dimensional assessments in practice. P5 noted that while they found these assessments compelling, they \textit{``probably wouldn't use it [as] a summative assessment,''} instead reserving them for more informal or formative contexts. Even when teachers valued the kinds of reasoning these assessments elicited, the time and labor required to grade them made it difficult to integrate such assessments into high-stakes settings.

Together, these concerns point to a key limitation in efforts to make data literacy an explicit assessment target. While three-dimensional assessments align well with instructional goals around scientific reasoning, they introduce tensions around both instructional focus and grading feasibility. This surfaces \textbf{Challenge 4: Balancing assessing data literacy and domain-specific learning goals}.
% , where making data literacy visible and assessable comes at the cost of practicality for teachers.

\section{Opportunities to Address Challenges in Assessing Data Literacy}
In this section, we discuss opportunities for addressing the four challenges identified in Section~\ref{sec:findings}. 

\subsection{Clarifying Conceptual Boundaries Between Data Visualization and Data Literacy}
Our findings in Section \ref{subsubsec:data_reasoning_practices} show that teachers often confuse data visualization with data literacy. Prior work frequently defines data literacy as an umbrella term involving practices that range from interpreting representations to making data-informed decisions \cite{Inverarity2022, Debruyne2021, Jiang2025}. Other work moves beyond these broader definitions to articulate the relationships between data literacy, data visualization, and related practices, such as Cruickshank et al.’s data education framework and Börner et al.’s data visualization literacy framework \cite{Cruickshank2024, borner2019}, however, these frameworks are rarely encountered by teachers and remain largely confined to research contexts. 

One opportunity to address this challenge is to support teachers through professional development and freely available lesson plans that make these distinctions explicit in practice. Professional development could center on teachers examining and revising their own assessment questions and discussing what kinds of student thinking each question is likely to elicit. By working through concrete examples rather than abstract definitions, teachers could more clearly distinguish between tasks that assess working with data, interpreting data representations, and applying domain knowledge. Lesson plans could complement this work by providing assessment examples that make these distinctions visible. For example, a lesson might include a set of related questions built around the same data, with annotations explaining which aspects of data literacy or domain knowledge each question is intended to assess. Seeing these distinctions in concrete classroom materials may help teachers clarify the conceptual boundaries between data visualization and data literacy.

\subsection{Making Real-World Data Easier to Adapt for K--12 Use}

Teachers also face challenges related to the data itself. As described in Section~\ref{subsubsec:data_source_choices}, even when teachers valued real-world data for its authenticity, they described altering it when they wanted to target specific misconceptions.

Real-world data is often incomplete and noisy, introducing challenges that complicate analysis and interpretation, particularly for novice learners \cite{Perini2024, Zhen2024}. At the same time, research in data literacy education emphasizes the pedagogical value of engaging with authentic data \cite{Ravi2024}. 
% Together, this work underscores that authenticity and instructional clarity are frequently in tension during assessment design.
Tools that support granular data cleaning and transformation, such as general-purpose data analysis and visualization platforms, can help manage real-world data but often introduce substantial overhead. These systems typically require teachers to learn complex interfaces and workflows, and they separate data sourcing from data cleaning and adaptation. As a result, the effort required to make real-world data usable could outweigh its benefits.

Some data literacy education approaches address this challenge by tightly coupling data generation, representation, and interpretation within a single, lightweight workflow. For example, Data Walking encourages learners to generate data through everyday actions and immediately represent and interpret it \cite{Hunter2025}. Such approaches illustrate how messiness can be made pedagogically productive when scope and complexity are appropriately managed. Therefore, rather than separating data sourcing, cleaning, and representation into distinct steps, approaches that lower the overhead of adapting real-world data while keeping those actions closely tied to assessment design may better align with teachers’ needs. One such direction is to support in-place data modification that allows teachers to make simple, targeted adjustments so real-world data can be aligned with instructional goals.

\subsection{Reducing Friction in Sourcing and Adapting Domain-Appropriate Visual Representations}

Teachers’ use of visualizations in assessments is often constrained by whether they can quickly locate or assemble a visualization that matches their domain conventions and quality expectations (Section \ref{subsubsec:pre_made_visuals}). 
% This challenge centers on the difficulty of finding the \emph{right visualization}---not just any representation, but one that is appropriate for a specific subject, level, and reasoning goal.

Several existing systems attempt to reduce this burden by centralizing access to data visualizations. For example, CODAP provides a set of standard visualization types that can be generated directly from datasets within a unified workspace \cite{CODAP}. Similarly, Gapminder offers a curated collection of datasets paired with a small number of well-defined visualization forms designed for exploratory analysis \cite{Gapminder}. These systems reduce the need for teachers to search the web by bundling data and representation in one place. However, the data and visualizations available in these tools may be insufficient. Teachers in our study described requiring highly specific visual forms (e.g., physics diagrams) that fall outside what is supported by these centralized tools. As a result, teachers often turn to ad hoc strategies such as image search, screenshots, or manual editing to obtain visual representations that meet their needs.

This suggests that reducing sourcing friction requires a larger body of visual representations that are organized by subject and task, rather than by chart type alone. One possible direction is to design systems where teachers can browse and reuse visuals contributed by other teachers, drawing from representations that have already been used in similar instructional contexts. Once teachers locate a visual that is close to what they need, they could make small adjustments, such as changing labels or values, without having to search across sites.

\subsection{Supporting Assessment of Data Literacy Alongside Domain Learning Goals}

Making data literacy and reasoning practices an explicit target of assessment (Section~\ref{subsubsec:3d_assessments}) creates a significant design challenge: it is difficult to simultaneously evaluate domain-specific knowledge and cross-cutting data literacy skills within a single task. This tension often forces teachers to choose between embedding data practices implicitly within disciplinary problems (Section~\ref{subsubsec:implicit_data_literacy}) or making them an explicit focus at the risk of de-emphasizing foundational content or increasing grading burdens (Section~\ref{subsubsec:feasibility_tensions}).

Prior work highlights the difficulty of assessing complex, open-ended reasoning in ways that are both interpretable and instructionally aligned. Systems such as CoGrader explore how human--AI collaboration can support evaluation of rich, project-based work, while other approaches combine automation and peer involvement to manage assessment complexity \cite{Chen2025, Liow2025}. However, these efforts primarily address how to assess open-ended work, rather than helping teachers decide what aspects of data literacy should be surfaced explicitly versus left implicit within disciplinary tasks.

One potential solution is to support structuring assessment questions so these goals are intentionally distinguished. For example, multiple-choice questions can include distractors that target different kinds of errors, some targeting misunderstandings of domain content and some targeting breakdowns in data reasoning, allowing teachers to assess both without relying on fully open-ended responses. Another similar approach is to hold the data representation constant while varying what is being assessed. For example, one part of a question might ask students to extract or describe patterns from a provided visualization, while a subsequent part asks them to apply domain-specific concepts to explain those patterns. This allows teachers to distinguish between data reasoning and disciplinary explanation without introducing separate tasks.

\section{Conclusion}

Across interviews with 13 teachers, we surfaced four recurring challenges that complicate efforts to make data literacy assessable in practice: (1) conceptual ambiguity between data visualization and data literacy, (2) tradeoffs between using real-world or synthetic data, (3) difficulty finding and adapting domain-appropriate visual representations and data visualizations, and (4) balancing assessing data literacy and domain-specific learning goals.

We also outlined several promising directions for addressing these challenges. First, teachers may benefit from clearer conceptual supports that distinguish visualization-specific skills from broader data reasoning practices, enabling more deliberate choices about what an assessment item is intended to elicit. Second, reducing friction in visualization sourcing will likely require expanding representational coverage beyond a narrow set of chart types and supporting lightweight adaptation, so that teachers can quickly locate or adjust materials that match disciplinary conventions and instructional intent. Third, supporting the use of authentic data may require lowering the overhead of adapting real-world datasets while keeping those actions close to instruction, allowing teachers to make targeted modifications that preserve key features of authenticity without introducing unnecessary complexity for students. Finally, approaches that clarify assessment intent and structure tasks to separate domain understanding and data reasoning may help teachers make data literacy explicit without having prohibitive grading costs.

Taken together, our work highlights assessment design as a productive lens for understanding how data literacy is enacted in classrooms and where support is most needed. By grounding solution directions in teachers’ lived constraints and practices, we aim to inform future efforts that help educators more consistently and feasibly assess students’ data literacy within the realities of K--12 teaching.

\bibliographystyle{ACM-Reference-Format}
\bibliography{workshop}

\appendix

\end{document}